\documentclass{svmult}
\usepackage{graphicx}

\usepackage[numbers]{natbib}
\usepackage{flafter}
\usepackage{csquotes}
\usepackage{epigraph}
\usepackage{epsfig}
\usepackage{color}
\usepackage{amssymb}
\usepackage[top=2.5cm, bottom=2.5cm, left=4cm, right=4cm]{geometry}
\usepackage[hidelinks]{hyperref}
\begin{document}
%  Custom commands, mostly for typesetting chemical formulae.
%\include{defcommands}

%  Load of stuff to coerce LaTeX into putting associated
%  floats in sensible places.
\newcommand\igorfig[1]{\includegraphics[scale=0.8]{#1}}
\renewcommand{\topfraction}{.9}
\renewcommand{\bottomfraction}{.3}
\renewcommand{\textfraction}{.1}
\renewcommand{\floatpagefraction}{.7}
\renewcommand{\dbltopfraction}{.75}
\renewcommand{\dblfloatpagefraction}{.85}
\setcounter{topnumber}{3}
\setcounter{bottomnumber}{2}
\setcounter{totalnumber}{20}
\setcounter{dbltopnumber}{3}
\setlength\epigraphwidth{11.5cm}
\setlength\beforeepigraphskip{0pt}
\setlength\epigraphrule{0pt}
\renewcommand{\epigraphsize}{\normalsize}
\setlength{\parindent}{15 pt}
%\setlength{\parskip}{11 pt}

%  More custom commands
\newcommand\mapfigure[3]{\begin{figure}
            \begin{center}
            \includegraphics[scale=0.8]{#1}
            \end{center}
            \caption{\label{#3}#2}
            \end{figure}
            }
%  Prohibit hyphenation for some acronyms
%  \hyphenation{NEXAFS PES HOMO LUMO}

% macro - prevents singletons at the top of pages
%\widowpenalty=10000 \clubpenalty=10000 \raggedbottom

%  Frontmatter
\title{Pauli's Principle in Probe Microscopy}
\author{SP Jarvis\inst{\dagger}, AM Sweetman\inst{\dagger}, L Kantorovich\inst{\ddagger}, E McGlynn\inst{\sharp}, and P Moriarty\inst{\dagger}}

\institute{\inst{\dagger}School of Physics and Astronomy, University of Nottingham, Nottingham NG7 2RD, UK \\ 
\inst{\ddagger} Department of Physics, King’s College
London, The Strand, London WC2R 2LS, UK\\
\inst{\sharp} School of Physical Sciences, Dublin City University, Glasnevin, Dublin 9, Ireland}

\maketitle

\epigraph{\flushright{\textit{It appears to be one of the few places in physics where there is a rule which can be stated very simply, but for which no one has found a simple and easy explanation. The explanation is deep down in relativistic quantum mechanics. This probably means that we do not have a complete understanding of the fundamental principle involved.}}{RP Feynman, \emph{The Feynman Lectures on Physics}, Vol III, Chapter 4 (1964)}}

\section*{Abstract}
Exceptionally clear images of intramolecular structure can be attained in dynamic force microscopy through the combination of a passivated tip apex and operation in what has become known as the ``Pauli exclusion regime'' of the tip-sample interaction. We discuss, from an experimentalist's perspective, a number of aspects of the exclusion principle which underpin this ability to achieve submolecular resolution. Our particular focus is on the origins, history, and interpretation of Pauli's principle in the context of interatomic and intermolecular interactions. \\

\begin{small}
{\noindent\textbf{Keywords:} dynamic force microscopy;non-contact atomic force microscopy; NC-AFM; Pauli exclusion principle; submolecular resolution; intramolecular; single molecule imaging}\\\\
\noindent
From \textit{Imaging and Manipulation of Adsorbates using Dynamic Force Microscopy},\\ Vol V of \textit{Advances in Atom and Single Molecule Machines}, Springer-Verlag;  \url{http://www.springer.com/series/10425}. To be published late 2014.

\end{small} 

\section{Intramolecular resolution via Pauli exclusion}
In 2009 the results of a pioneering dynamic force microscopy (DFM\footnote{Although the term non-contact atomic force microscopy (NC-AFM) is widespread -- to the extent that the major conference in the field is the annual International NC-AFM meeting -- it is arguably something of a misnomer to label the technique ``non-contact'' when it is now commonplace to operate in a regime where the probe is in contact with the sample. We will therefore use the term \textit{dynamic} force microscopy throughout this chapter.}) experiment by Leo Gross and co-workers at IBM Z\"urich were published\cite{Gross:2009} and revolutionised the field of scanning probe microscopy. Gross \emph {et al.} captured arguably the clearest real space images of a molecule achieved up to that point, resolving the ``textbook'' structure of the molecular architecture. Two important experimental protocols enabled Gross \emph {et al.} -- and, subsequently, a number of other groups\cite{deOteyza:2013,Zhang:2013,Sweetman:2014,Hapala:2014,Pavlicek:2012,Pawlak:2011,Riss:2014} (see Fig. 1.1 for examples) -- to attain this exceptionally high resolution. First, the apex of the probe was functionalised (by picking up a molecule) to render it inert. This enabled the scanning probe to be placed extremely close to the adsorbed molecule of interest -- so close that the second experimental protocol, namely the exploitation of electron repulsion via the Pauli exclusion principle\footnote{We shall return, in Sections 1.4 and 1.5, to a detailed discussion of whether or not it is appropriate to describe the effects of Pauli exclusion as a repulsive force.}, played a key role in the imaging mechanism.

It is this second protocol which is the primary focus of this chapter. We'll discuss just how Pauli exclusion is exploited in state-of-the-art scanning probe microscopy, what pitfalls there might be in interpreting features in DFM images as arising directly from chemical bonds, and to what extent scanning probe measurements of tip-sample interactions provide deeper experimental insights into the exclusion principle itself. We should also stress right from the outset that although we concentrate on dynamic force microscopy throughout this chapter, prior to Gross \emph {et al.}'s 2009 paper, Temirov, Tautz and co-workers had achieved unprecedented spatial resolution using a technique for which they coined the term scanning tunnelling hydrogen microscopy (STHM)\cite{STHM1,STHM2,STHM3,STHM4}. Both STHM and the type of DFM imaging introduced by Gross \emph {et al.}\cite{Gross:2009} exploit Pauli exclusion as a means to acquire exceptionally high resolution. Before covering the exploitation of the exclusion principle in scanning probe microscopy, we'll consider a number of aspects of the fascinating history of Pauli's \textit{Ausschlie{\ss}ungsregel}\cite{Massimi} and outline some of the rich physics underpinning the principle.

\begin{figure}
\centering
\includegraphics[width=0.9\linewidth]{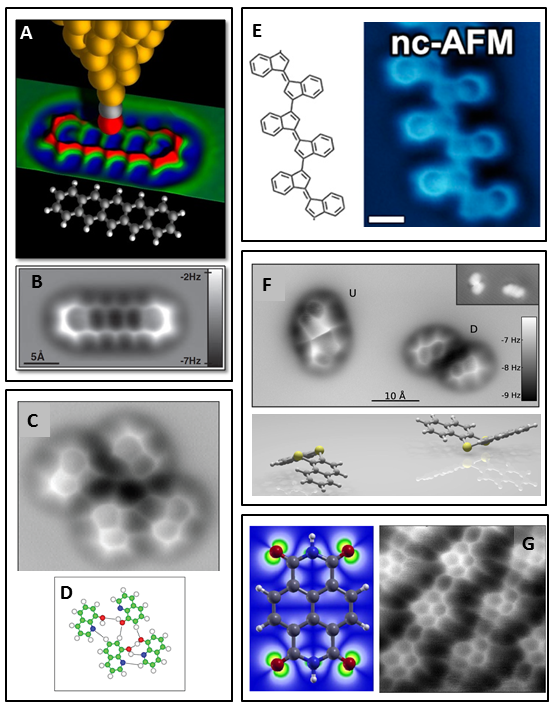}
\caption{\textbf{Imaging bonds via the Pauli exclusion principle}.\textbf{(A)} Combination of schematic illustration and experimental data to demonstrate experimental protocol used to acquire submolecular resolution. The apex of the probe used in a dynamic force microscope is passivated (in this case with a CO molecule) and scanned across a pentacene molecule at a height where Pauli exclusion plays a key role in determining the tip-sample interaction. \textbf{(B)} Experimental frequency shift image for a pentacene molecule. [A and B taken from Gross et al.\cite{Gross:2009}. \copyright American Assocation for the Advancement of Science (2009)]. \textbf{(C)} Dynamic force microscope image of four 8-hydroxyquinoline molecules. Both intra- and intermolecular features are observed. (See Section 1.7). \textbf{(D)} Schematic diagram of molecular arrangement shown in (C) with the expected positions of hydrogen-bonds drawn as lines between the molecules. [C and D taken from Zhang et al.\cite{Zhang:2013}. \copyright American Assocation for the Advancement of Science (2012)]. \textbf{(E)} High resolution image of a chain of oligo-(E)-1,1′-bi(indenylidene) with associated structural model. Taken from Riss \emph{et al.}\cite{Riss:2014}.\copyright American Chemical Society (2014). \textbf{(F)} DFM image of two different conformers of dibenzo[a,h]thianthrene on a NaCl/Cu(111) substrate with (lower panel) structural models of both conformers. Taken from Pavlicek et al.\cite{Pavlicek:2012}.\copyright American Physical Society (2012).\textbf{(G)} Structural model of a naphthalenetetracarboxylic diimide (NTCDI) molecule and a DFM image of a hydrogen-bonded assembly of NTCDI molecules. From Sweetman et al.\cite{Sweetman:2014}.\copyright Nature Publishing Group (2014).}
\label{fig:Fig1}
\end{figure}

\section{A potted history of Pauli's exclusion principle}
Michela Massimi has written an authoritative and engaging history of the Pauli exclusion principle (PEP)\cite{Massimi}, which impressively combines clear explanations of the quantum and statistical physics underlying the PEP with engaging discussions of both the history and the philosophical ramifications of the principle. As Massimi points out in the preface to her book, her research on the origin and validation of the exclusion principle took almost ten years. For those readers interested in a comprehensive account of the ``evolution'' of the PEP we therefore strongly recommend Masimi's book. Here we will limit ourselves to providing a brief summary of those aspects of the PEP which are of key significance for (ultra)high resolution scanning probe microscopy.

The origins of the exclusion principle lie, like so many aspects of quantum physics, in the interpretation of spectroscopic data. In particular, a series of so-called anomalies in the spectra of alkali and alkaline earth metals, and, arguably more importantly, the response of atomic spectra to the application of a magnetic field, i.e. the (``anomalous'') Zeeman effect, became a major challenge to the Bohr-Sommerfeld theory of the electronic structure of atoms in the early 1920s. It was only with the introduction of what came to be known as electron spin -- but which Pauli initially called simply the electron \textit{Zweideutigkeit} (``twofoldness'') -- that the spectroscopic data could be reconciled with the theoretical predictions. The introduction of electron \textit{Zweideutigkeit}\cite{Pauli:1925a} was followed very closely by Pauli's statement of the exclusion principle\cite{Pauli:1925b} (or, as it was known at the time, the exclusion rule). Pauli subsequently won the Nobel prize in 1945 for his discovery of the exclusion principle.

It is worth quoting directly from Pauli's Nobel lecture, given on Dec. 13 1946, as this provides key insights into the original formulation of the principle ``straight from the horse's mouth'', as it were:

\begin{quote}

On the basis of my earlier results on the classification of spectral terms in a strong magnetic field the general formulation of the exclusion principle became clear to me. The fundamental idea can be stated in the following way: 

The complicated numbers of electrons in closed subgroups are reduced to the simple number \emph{one} if the division of the groups by giving the values of the four quantum numbers of an electron is carried so far that every degeneracy is removed. An entirely non-degenerate energy level is already \emph{closed}, if it is occupied by a single electron; states in contradiction with this postulate have to be excluded.

\end{quote}
\noindent
Or, if we couch this in the lexicon of modern quantum mechanics, no two electrons can have the same values of $n$, $l$, $m_l$, and $m_s$ (i.e. the principal, orbital angular momentum, magnetic, and spin quantum numbers). More succinctly, no two electrons can occupy the same quantum state. (The Pauli exclusion principle of course holds for all fermions (half-integer spin particles), not just electrons. We'll return to this point very soon).

Pauli's \emph{Zweideutigkeit} is now of course known as particle spin but the inferred connection with the classical concept of a spinning object is unfortunately misleading. Indeed, Pauli himself switched from being firmly opposed to any connection between his \emph{Zweideutigkeit} and spin, to a somewhat grudging acceptance of a link, and then, as his Nobel lecture highlights, back to a significant degree of scepticism about the value of any classical analogy:

\begin{quote}

On the other hand, my earlier doubts as well as the cautious expression ``classically non-describable two-valuedness'' experienced a certain verification during later developments, since Bohr was able to show on the basis of wave mechanics that the electron spin ... must therefore be considered as an essentially quantum-mechanical property of the electron.

\end{quote}

\subsection{Particle statistics and the quantum identity crisis}

Following hot on the heels of Pauli's publication of the exclusion principle, first Fermi\cite{Fermi:1926,FermiTranslation} and then Dirac\cite{Dirac:1926} explored the quantum statistics of an ideal gas of particles which was subject to the exclusion principle. Dirac coined the term \emph{fermion} to describe a particle subject to the Fermi-Dirac statistics he and Fermi derived; a fermion is therefore a particle which obeys the Pauli exclusion principle (and concomitantly is of half-integer spin). At the very heart of quantum statistics -- and, indeed, of classical statistical mechanics -- lies the issue of the distinguishability of particles\footnote{Long before the advent of quantum mechanics, the effect of considering indistinguishable vs distinguishable particles on the partition function for a system was known as the Gibbs paradox in classical thermodynamics/statistical mechanics}. A simple back-of-the-envelope argument based on the (in)distinguishability of particles can provide a helpful insight into the origin of the exclusion principle\cite{DaviesTextbook}.

Before we introduce that back-of-the-envelope approach, however, it is first important to define just what it is we mean by indistinguishable particles. This, despite first appearances, is a far from trivial question to address and has been the subject of quite considerable debate and interest for many decades. De Muynck\cite{DeMuynck:1975}, Berry and Robbins\cite{Berry-Robbins:2000}, Ginsberg \emph {et al.}\cite{Ginsberg:2007} (see also Fleischhauer\cite{Fleischhauer:2007} for a very readable overview of Ginsberg et al's work), Omar\cite{Omar:2005}, and Dieks and co-workers\cite{Dieks:2008, Dieks:2011}, amongst many others, have considered and explored the important issue of how indistinguishability and quantum statistics are intrinsically coupled. We shall not delve into the detailed arguments -- be they physical, philosophical, or semantic in scope -- and instead restrict ourselves to the following relatively simple, although certainly not ``universal'', definitions. (It is also important to note that the condition for antisymmetry and the exclusion principle are not equivalent statements).

First, we draw a distinction between \emph{identical} and \emph{indistinguishable} particles. Identical particles are those which have the same intrinsic (or ``internal'') properties (and the same values associated with those intrinsic properties), i.e. mass, charge, spin. So two electrons are identical to each other. And two protons, or two neutrons, are similarly identical to each other. But electrons are clearly not identical to protons, nor to neutrons. (We apologise for labouring the point to this extent but the terms ``identical'' and ``indistiguishable'' are often used interchangeably -- including in many textbooks -- and this has led to quite some confusion at times).

If we have a collection of identical particles then they are \emph{indistinguishable} if we cannot separate them on the basis of their ``external'' properties such as position or momentum. But classically it \emph{is} possible to distinguish between identical particles (at least in principle): we can effectively ``label'' individual identical particles on the basis of their positions or trajectories and distinguish them accordingly\footnote{In a thought-provoking paper, Versteegh and Dieks\cite{VersteeghAJP} discuss the importance of the distinguishability of identical particles and what these means for classical thermodynamics and statistical mechanics, including the Gibbs paradox. We note, however, that there is a very important omission in the list of papers cited by Versteegh and Dieks, namely a paper by Edwin Jaynes\cite{Jaynes} who makes the point, following a similar analysis by Pauli, that the classical thermodynamic definition of entropy as the integration of d$Q$/$T$ over a reversible path is only introduced in the context of constant particle number. This means that there is always (ultimately, see Ehrenfest and Trkal\cite{Ehrenfest}) an arbitrary integration function (not an integration constant, but a function of $N$) that can be used to yield the desired extensivity of the entropy.}. Quantum mechanically, however, the standard argument is that due to delocalisation we lose this ability to label particles on the basis of their trajectories and they then become indistinguishable. 

But to what extent is this true? Are quantum particles indeed indistinguishable? One can find undergraduate-level descriptions of quantum statistics\cite{Rohlf} which claim that quantum particles can in fact be distinguished on the basis of what might be called a ``Rayleigh criterion'' for wavepackets: if two particles are separated by a distance greater than their de Broglie wavelength (i.e. such that the wavefunction overlap is minimal) then they are distinguishable on the basis of their respective positions. Versteegh and Dieks\cite{VersteeghAJP} invoke similar arguments about the spatial extent of wavepackets enabling identical quantum particles to be distinguished.

However, whether this is a valid condition for distinguishability is far from clear-cut. In his commentary on Ginsberg \emph {et al.}'s work\cite{Ginsberg:2007}, Fleicschhauer\cite{Fleischhauer:2007} states the following:

\begin{quote}
In the quantum world, particles of the same kind are indistinguishable: the wavefunction that describes them is a superposition of every single particle of that kind occupying every allowed state. Strictly speaking, this means that we can't talk, for instance, about an electron on Earth without mentioning all the electrons on the Moon in the same breath.
\end{quote}

Why might Fleicschhauer say this?\footnote{It is perhaps worth noting at this point that the ``interconnectedness'' to which Fleicschhauer alludes in this quote, and its relevance (or not) to the Pauli exclusion principle, was the subject of a great deal of sometimes ill-tempered online debate following the BBC's broadcast of a popular science lecture on quantum mechanics by Brian Cox, which included a discussion of the PEP. Jon Butterworth's post for The Guardian\cite{Butterworth} is a short, clear and entertaining discussion of the furore and the physics surrounding Cox's lecture.} The answer is, from one perspective at least, rather straight-forward. The universal superposition to which Fleicschhauer refers arises because in reality we never have perfect confinement of particles: there is no such thing as the infinite potential well beloved of introductory quantum physics courses and there is therefore some finite (albeit extremely small) probability for tunnelling. Thus, in this sense an electron on the Earth is indeed indistinguishable from an electron on the Moon (or on Alpha Centauri).

But what really matters, of course, are the effects that this type of ``coupling'' might have on experimental measurements. And for electrons separated by centimetres, let alone light years, those effects are, to put it mildly, utterly negligible. If we consider a "double well" system for an electron on Earth and an electron on Alpha Centauri, the energy level splitting is unimaginably tiny (and beyond anything we could ever begin to hope to measure), and the time-scale for evolution of the quantum state exceeds the age of the universe.  

So \emph{in any practical sense}, position can indeed be used to distinguish quantum particles. This is why we can treat electrons in well-separated atoms as being distinguishable. In principle, the electrons are indeed described by a single multi-particle (``universal'')  wavefunction and are thus indistinguishable. In practice, however, the spatial extent of the particle wavepacket is such that we can treat electrons in atoms separated by distances much greater than their equilibrium bond length as distinguishable. Only when those atoms are brought together so that there is appreciable overlap of electronic wavefuctions, as in chemical bond formation or, as we shall discuss below, a dynamic force microscopy experiment, can one state that the electrons on each atom become indistinguishable.

Following this lengthy ``detour'' on the topic of distinguishability vs indistinguishability, we are now finally at the point where we can return to a consideration of that back-of-an-envelope argument for the PEP which was mentioned above.

\section{Statistics, symmetry, and spin} 
Let's take a system where identical quantum particles can't be distinguished from another. As the particles are indistinguishable then when we compute the probability density for the system, i.e. $|\Psi|^2$, we must get the same answer regardless of how we arrange the particles, i.e. their spatial positions have no influence on the probability density. We'll consider a very simple system with just two particles whose positions are $\textbf{r}_1$ and $\textbf{r}_2$ and whose single particle wavefunctions are $\psi_1$ and $\psi_2$ respectively. If we cannot distinguish Particle 1 from Particle 2 then it's clear that
\begin{equation}
|\Psi(\textbf{r}_1,\textbf{r}_2)|^2 = |\Psi(\textbf{r}_2,\textbf{r}_1)|^2 
\end{equation}

\noindent
This means one of two things. Either

\begin{equation}
\Psi(\textbf{r}_1,\textbf{r}_2) = \Psi(\textbf{r}_2,\textbf{r}_1)
\end{equation}
\noindent
or
\begin{equation}
\Psi(\textbf{r}_1,\textbf{r}_2) = -\Psi(\textbf{r}_2,\textbf{r}_1)
\end{equation}

To meet the condition imposed by Eqn. 1.2, we must have the following two-particle wavefunction:

\begin{equation}
\Psi(\textbf{r}_1,\textbf{r}_2) = \frac{1}{\sqrt{2}}\big(\psi_1(\textbf{r}_1)\psi_2(\textbf{r}_2)+\psi_2(\textbf{r}_1)\psi_1(\textbf{r}_2)\big)
\end{equation}

\noindent
Or to satisfy Eqn. 1.3 we need the following:

\begin{equation}
\Psi(\textbf{r}_1,\textbf{r}_2) = \frac{1}{\sqrt{2}}\big(\psi_1(\textbf{r}_1)\psi_2(\textbf{r}_2)-\psi_2(\textbf{r}_1)\psi_1(\textbf{r}_2)\big)
\end{equation}

Eqn. 1.4 represents what is called the symmetric case, while Egn. 1.5 is termed the antisymmetric case \footnote{The use of the terms symmetric and antisymmetric follows from Eqn. 1.2 (where $\Psi$ is a symmetric function with respect to the exchange of coordinates) and Eqn 1.3 (where $\Psi$ is an antisymmetric function). Note also that the factor of $\frac{1}{\sqrt{2}}$ in Eqn. 1.4 and Eqn. 1.5 arises from normalisation of the wavefunction}. The antisymmetric equation leads us to a simple, but exceptionally important, result -- a result that is at the very core of how the universe behaves because it is ultimately responsible for the stability of matter\cite{Lieb:1975,Lieb:1976,Lieb:1990}. Note what happens when we make $\psi_1 = \psi_2$ in Eqn. 1.5 (or, in other words, we put both particles in the same quantum state): \emph{the two-particle wavefunction, $\Psi$, vanishes}. \emph{This} is the essence of the Pauli exclusion principle: in the antisymmetric case, no two particles can exist in the same quantum state\footnote{We are neglecting explicit consideration of the spin contribution here -- see Section 1.3.1. Moreover, we are making drastic simplifications regarding the treatment of many electron systems in order to put across the ``essence'' of the exclusion principle. For example, equations 1.4 and 1.5 are approximations because, in reality, there are many more contributing terms (as in the Configuration Interaction method of quantum chemistry. See Kantorovich\cite{KantorovichBook} for a summary.)}. (We should also stress that the exclusion principle is\textit{ not equivalent} to the statement that fermions have antisymmetric wave functions. Rather, the exclusion principle follows from the antisymmetric character of fermions).

A rather remarkable observation is that \emph{only} antisymmetric and symmetric wavefunctions are found in nature for fundamental particles, i.e. we only have bosons (symmetric state) and fermions (anti-symmetry).  No other particles have been found that fall outside these symmetry classes\footnote{Note, however, that the key principle underlying the concept of \emph{supersymmetry} is that bosons can be converted into fermions and vice versa. Supersymmetry therefore introduces a bosonic partner for every fermion (and, again, vice versa). To the chagrin of (some of) the particle physics community, however, any evidence for supersymmetry remains frustratingly elusive. Moreover, we are omitting any discussion of quasiparticles here. The results of measurements of two-dimensional systems exhibiting the fractional quantum Hall effect have been interpreted in terms of anyons\cite{Wilczek:1982}, quasiparticles with mixed symmetry.} As Omar\cite{Omar:2005} points out in a comprehensive and very readable review of the ramifications of indistinguishability in quantum mechanics, this existence of only symmetric and antisymmetric states\footnote{...for the \emph{total} wavefunction. Again, see Section 1.3.1.} is best described as a postulate (the ``symmetrization postulate''). And, disconcertingly, it's a postulate that apparently can't be deduced from the framework of quantum mechanics (either the non-relativistic or relativistic ``breeds'' of the theory). In other words, we simply have to accept that only bosons and fermions exist (or, at least, we have no good experimental evidence to date for fundamental particles arising from other rather more exotic statistics/symmetries such as parastatistics (see Omar\cite{Omar:2005})). In this sense, we have progressed very little since Pauli voiced his misgivings about the origin of the exclusion principle almost seventy years ago:

\begin{quote}
I was unable to give a logical reason for the Exclusion Principle or to deduce it from more general assumptions... in the beginning I hoped that the new quantum mechanics would also rigorously deduce the Exclusion Principle.
\end{quote}

\subsection{Putting a spin on the story}
All known fundamental particles are either bosons or fermions. (Within the Standard Model, fermions are ``matter'' particles, whereas bosons are generally force ``carriers''\footnote{...although the Higgs boson is an honourable exception.}. Again, we are not including quasiparticles in the discussion.). All bosons have integer spin while fermions have half-integer spin. Clearly there must be a strong connection between spin and symmetry. Indeed, this is known as the spin-statistics theorem and holds not just for individual particles but composites of fundamental particles. 

This link between spin, statistics, and the exclusion principle, however, very much appears not to be something that can be deduced from non-relativistic quantum mechanics. This is the origin of the statement from Feynman quoted at the start of this chapter -- the link between spin and the exclusion principle is ``deep down'' in relativistic quantum mechanics. More recently, Bartalucci \emph {et al.}\cite{Bartalucci:2006} have put it like this:

\begin{quote}
Although the principle has been spectacularly confirmed by the number and accuracy of its predictions, its foundation lies deep in the structure of quantum field theory and has defied all attempts to produce a simple proof...
\end{quote} 

\noindent This means that within the non-relativistic quantum framework the spin-statistics-symmetry link is generally accepted as a dictum, although alternative non-relativistic approaches have certainly been explored\cite{Berry-Robbins:2000}. (Duck and Sudarshan\cite{DuckSudarshan} detail a proof of the spin-statistics theorem which can be ``recast'' in non-relativistic quantum field theory, but only if an aspect of their proof which stems from relativistic quantum theory (via Lorentz invariance) can be invoked as a postulate).

Notwithstanding its essential relativistic origin, the spin contribution can be incorporated into the particle wavefunction in non-relativistic quantum mechanics in a straight-forward fashion via the introduction of the spin orbital. A spin orbital is a product of a spatial wavefunction (such as those described in the preceding section) and a spin function, which we can represent as $\chi(\uparrow)$ or $\chi(\downarrow)$ for the spin-up and spin-down states respectively. So, if we use $\textbf{x}$ as a variable which incorporates both the spatial and spin coordinates, and we switch to using $\phi$ to represent only the spatial part (so that we can, as per convention, use $\psi$ to represent the wavefunction), we have the following for the spin-up state of an electron:

\begin{equation}
\psi(\textbf{x}_1)=\phi(\textbf{r}_1) \chi(\uparrow)
\end{equation}  

\noindent We therefore now have two options for ensuring antisymmetry in a two electron (or multi-electron) system: either the spatial part \textit{or} the spin part can lead to an antisymmetric total wavefunction, $\Psi(\textbf{x}_1,\textbf{x}_2)$. In other words, if two electrons have opposite spin states then there is no constraint on the spatial wavefunction. But this is nothing more than the statement of the Pauli exclusion principle given earlier: no two electrons can exist in the same quantum state.

\section{The origin of Pauli repulsion: A Gedankenexperiment} 
At short interatomic or intermolecular separations, Pauli repulsion\footnote{We focus throughout this chapter only on fermions. For bosons, and as discussed by Mullin and Blaylock\cite{MullinBlaylock}, an effective \textit{attractive} force is often invoked.} is much stronger than any electrostatic interaction, increasing \emph{very} rapidly with decreasing distance between atoms or molecules.  Recall, for example, that the Pauli repulsion term in the Lennard-Jones potential is modelled not with a $\frac{1}{r}$ dependence, as one would expect for a classical electrostatic interaction (between point charges), but with a $\frac{1}{r^{12}}$ function. This $\frac{1}{r^{12}}$ dependence is, of course, purely empirical in the Lennard-Jones (L-J) potential -- it has no grounding in theory -- but, nonetheless, the exceptionally high sensitivity of the repulsive interaction to small changes in interatomic/intermolecular separation is captured well by the functional form. 

Of course, and as Baerends\cite{Baerends:1992} discusses in a clear overview of Pauli repulsion effects in adsorption, we are dealing not with point charges and a pure Coulombic interaction but with a screened Coulomb potential and delocalised electron ``clouds''. The overlap of the electron clouds at short separations leads in a classical model, and perhaps counter-intuitively, to an \emph{attractive} electrostatic interaction. It is only when the inter-atomic separation becomes so small that nuclear repulsion dominates that the overall electrostatic force becomes repulsive.

Thus, and as we hope is abundantly clear from previous sections, we cannot expect to understand electron repulsion due to Pauli exclusion in the context of classical electrostatics. The fundamental origin of the repulsion comes from, as we've seen, the physical impossibility of ``squeezing'' two fermions into the same quantum state. But the central question is this: just how does the exclusion principle translate into a physically measurable interaction? We'll see in the following section how dynamic force microscopy allows us to directly probe the exclusion-derived repulsion between the electron density of two atoms or molecules. Before we consider the results of the real-world experiment, however, it's very helpful to think about a ``stripped-down'' system involving the overlap of two single particle wavefunctions (see Section 1.3)\cite{WilsonGoddard1,WilsonGoddard2,WilsonGoddard3}. This ``\textit{Gedankenexperiment}'', if you will, provides compelling insights into the origin of Pauli repulsion.

First, recall that the kinetic energy operator is $-\frac{\hbar^2}{2m}\nabla^2$. The curvature of a wavefunction therefore determines its kinetic energy (via the Laplacian, $\nabla^2$).  Wilson and Goddard's approach\cite{WilsonGoddard1} to elucidating the origin of Pauli repulsion was to compare the kinetic energy (KE) of a Hartree product of the wavefunctions for two same-spin electrons with the KE of an antisymmetrized product (see Fig. 1.2). A Hartree product is simply the following:

\begin{equation}
\Psi_{\textrm{\small{Hart}}}(\textbf{r}_1,\textbf{r}_2) = \psi(\textbf{r}_1)\psi(\textbf{r}_2)
\end{equation}   

\noindent As should be clear from Section 1.3, the multiparticle wavefunction $\Psi_{\textrm{\small{Hart}}}$ is not antisymmetric (nor does it take into account indistinguishability of the particles) and is therefore in general not appropriate to use to describe fermions. However, we can take the Hartree product as a representation of the system when the Pauli exclusion principle is ``suppressed'' and determine the resulting kinetic energy.

  \begin{figure}[b!]
  \centering
  \includegraphics[width=0.75\linewidth]{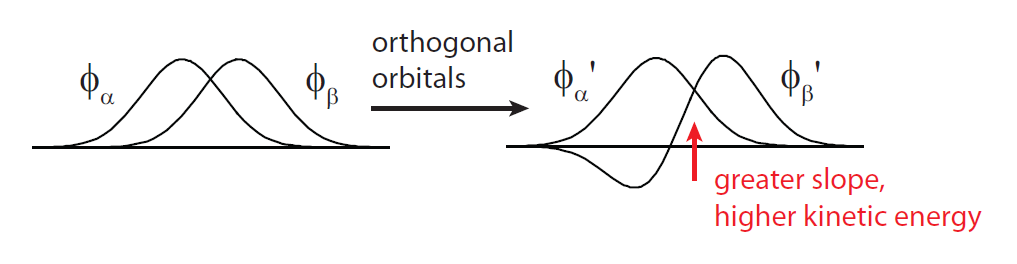}
  \caption{The effective repulsion due to Pauli exclusion stems from the change in the curvature of the wavefunction due to the requirement for antisymmetrization in fermion systems. One approach to visualising this is to consider the orthogonalization of orbitals (which is placed as a constraint on Slater determinant approaches to constructing a multi-particle wavefunction). Higher wavefunction curvature leads to a higher kinetic energy. Equivalently, higher curvature is accounted for in Fourier space by higher spatial frequency (momentum) components. Figure taken from the PhD thesis of Julian Su\cite{SuThesis}. \copyright Julian Su (2007).}
  \label{fig:Fig2}
  \end{figure}

In order to incorporate Pauli exclusion we have to consider a multi-particle wavefunction which is appropriately\textit{ anti-symmetrized}. Slater introduced an elegant method of enforcing this antisymmetry requirement via the determinant approach which now bears his name\cite{Slater:1929}. Wilson and Goddard\cite{WilsonGoddard1} focussed on the orthogonality of orbitals which is generally \textit{imposed} in approaches which treat the multiparticle wavefunction in terms of (a sum of) Slater determinants (see Fig. 1.2, taken from the PhD thesis of Julius Su\cite{SuThesis}). We note, however, that orthogonality is a constraint on the multiparticle wavefunction \textit{that is not strictly necessary}\cite{KantorovichBook} and, as discussed by Beylkin et al.\cite{Beylkin:2008} leads to ever-increasing levels of computational expense as the size of a system grows.

Nonetheless, to ensure antisymmetry (i.e. the requirement of Eqn. 1.5), wavefunction slope and curvature must necessarily increase and thus the overall picture emerging from Fig. 1.2 is correct (even if one doesn't invoke orthogonality as the root cause of the increase in wavefunction curvature). This change in curvature results in a corresponding increase in kinetic energy. A complementary explanation from a Fourier analysis perspective, as noted in the following section, is that the increase in curvature of the wavefunction necessitates the introduction of higher spatial frequency contributions, i.e. higher \emph{momentum} components). It is this increase in KE (or momentum) which is responsible for the majority of Pauli repulsion.

There are two important assumptions built into this description of Pauli exclusion, however. First, we have adopted a ``pairwise'' approach to considering electron-electron interactions when, in reality, Pauli exclusion is an \emph{n}-body, rather than a two-body problem. The second, and related, issue is that the modification of the wavefunction due to orthogonalisation will mean that the electron density will be distributed differently, affecting electron-electron interactions and giving rise to the effect known as correlation. Interactions between same-spin electrons go by the name \textit{Fermi correlation}, whereas those between opposite-spin electrons are known as \textit{Coulomb correlation}\footnote{The combined contributions of the exclusion principle and electron correlation produce the exchange-correlation contribution to the functional in density functional theory.}. Nonetheless, the dominating contribution to Pauli repulsion is the pure quantum-mechanical component arising from wavefunction antisymmetry.

\section{Is there a Pauli exclusion \emph{force}?}
Having spent much of the chapter up to this point using the term ``Pauli repulsion'', it might seem a little perverse for us to now pose the question as to whether there is a Pauli exclusion force or not (particularly as the experimental technique we're considering is dynamic \emph{force} microscopy). Notwithstanding the use of ``\textit{Pauli repulsion}'' or ``\textit{Pauli exclusion force}'' in the DFM literature -- and, more broadly, throughout very many areas of science (spanning, for example, particle physics, single molecule imaging and spectroscopy, astrophysics\footnote{The Pauli exclusion principle prevents the collapse of white dwarf and neutron stars. See \textit{Neutron Stars 1: Equation of State and Structure}, P. H\"ansel, AY Potekhin, and DG Yakovlev, Springer (New York, 2007).}, and cosmology) -- a number of authors have made the claim that Pauli exclusion does not produce a force in the traditional sense. Mullin and Blaylock\cite{MullinBlaylock}, in particular, present a set of arguments as to why they are of the opinion that couching the effects of Pauli exclusion in terms of a repulsive \textit{force}, or exchange \textit{force}, can be rather misleading. Indeed, they go so far as to argue -- and we quote directly from their paper -- that \emph{``there is no real force due to Fermi/Bose symmetries''}, citing, amongst others, Griffiths' description of the effects of Pauli exclusion\cite{Griffiths}:

\begin{quote}
We call it an exchange force but it is not really a force at all - no physical agency is pushing on the particles; rather it is purely a geometric consequence of the symmetrization requirement. 
\end{quote}

\noindent What does Griffiths (and, by extension, Mullin and Blaycock) mean by this? 

To back up their assertion that Pauli ``repulsion'' is not a force in the traditional sense, Mullin and Blaycock's consider a number of ``archetypal'' physicochemical phenomena where the exclusion principle plays a key role. Arguably the most instructive of these is their discussion of the changes in momentum in a classical gas as compared to a Fermi gas. We encourage the reader to follow the detail of the analysis in Section II of their paper (under the sub-section entitled \textit{Virial Expansion}) and restrict ourselves here simply to highlighting the central point they make. 

Consider first a classical ideal gas in a container. Pressure, $P$, arises from the combined impacts of each atom of that gas on the walls of the container and is given by the force per unit area. Force, in turn, is the rate of change of momentum. The mean force, $\bar{F}$, which each individual molecule of the gas contributes is $\bar{F}={\Delta p}/{\Delta t}$, where $\Delta p$ is the momentum change on striking the wall. (This is twice the atomic momentum because the sign of the momentum flips on collision). $\Delta t$ is the time required for an atom to cross the container, i.e. $\Delta t = mL/\bar{p}$ where $L$ is the width of the container and $m$ is the atomic mass. The key point in the classical case is this: if we make the volume of an ideal gas smaller or we introduce repulsive interactions (with no change in temperature), the pressure of the gas will rise because of a decreased $\Delta t$ due to a change in (the effective) $L$ arising from collisions, \emph {but $\bar{p}$ remains the same}. (Recall that for a classical gas the root mean square momentum, $p_{rms}$ is $\sqrt{3mk_BT}$) 

Compare this to what happens for a Fermi gas subject to the  exclusion principle. The effect of the exclusion principle is to modify the \textit{momentum distribution}. Mullin and Blaylock argue that this is subtly different to what happens for the classical gas when repulsive interactions are introduced. Classically, the repulsive forces raise the pressure of the gas because the collisions and deflections of the atoms change the atomic transit time. Quantum-mechanically, the momentum distribution is ``intrisically'' modified because of the higher curvature of the wavefunction which results from the exclusion principle. Position and momentum are conjugate variables and are thus two sides of the same coin - Fourier transformation allows us to switch between the two (entirely equivalent) representations. The higher wavefunction curvature demanded by Pauli exclusion is entirely equivalent to stating that higher spatial frequency components are required in reciprocal (i.e. momentum) space\footnote{This, of course, is the fundamental origin of the Heisenberg uncertainty principle.}. It is this intrinsic symmetry-driven modification of the momentum distribution which raises the pressure of the Fermi gas.

It is worth lifting another couple of quotes from Mullin and Blaylock's paper to highlight just how strongly opposed they are to equating Pauli exclusion with a repulsive force:

\begin{quote}
The idea of an effective repulsion between fermions ignores the real physics and gives a very poor analogy with classical repulsive gases...we offer the following guiding principle regarding statistical symmetries: ``May the force be \emph{not} with you''.
\end{quote}
 
\noindent Is this degree of anti-force scepticism justified, however?

\section{Beyond Gedanken: Exploiting exclusion in force microscopy}
 
At this point, the pragmatic scanning probe microscopist could quite reasonably take issue with the preceding arguments because the primary experimental observable in a dynamic force microscopy experiment is the frequency shift of the probe. And this, via the Sader-Jarvis formalism\cite{Sader:2004}, for example, can be converted directly to a tip-sample \textit{force}. The effects of Pauli exclusion are directly measurable in DFM because they shift the resonant frequency of the probe-cantilever system, and this ultimately can be interpreted as a change in the tip-sample force. Notwithstanding the arguments put forward by Mullin and Blaylock\cite{MullinBlaylock}, and Griffiths\cite{Griffiths}, amongst others, if Pauli exclusion isn't giving rise to a force then it certainly very much looks like it in a DFM experiment.
 
The resolution of this apparent conflict may lie, as Moll \emph {et al.} have discussed in a recent paper focussed on the interpretation of submolecular resolution DFM images\cite{Moll:2012}, in the virial theorem. Slater showed in the 1930s that the virial theorem can be applied to a molecule\cite{SlaterVirial}, assuming that the nuclei are fixed in place by external forces. The total electron energy, $E$, is related to the electronic kinetic energy, $T$, and potential energy, $V$, as follows:
 
 \begin{eqnarray}
 T=-E-r(\frac{dE}{dr})\\
 V=2E+r(\frac{dE}{dr})
 \end{eqnarray}
 
 The electronic kinetic energy and potential energy are thus coupled via the virial theorem. Moll \emph {et al.}\cite{Moll:2012} claim that, despite the Pauli exclusion force being non-conservative in character, if it is assumed that we have a diatomic (or dimolecular) system with one degree of freedom -- as is the case for the tip-sample system in DFM -- the Pauli energy and the increase in electronic kinetic energy can be related as follows:
 
 \begin{equation}
 E_{\textrm{\small{Pauli}}}(z)=\frac{1}{z}  \int_z^{\infty}\Delta E_{kin}(z')dz'
 \end{equation}
 
\noindent where $z$ is the interatomic/intermolecular separation. The issue of extracting accurate measures of non-conservative forces from the frequency shift observable in DFM, however, continues to attract considerable debate and discussion. For example, the Sader-Jarvis inversion technique\cite{Sader:2004} widely applied to extract forces from frequency-shift-vs-separation curves must, as John Sader and his co-authors themselves highlight\cite{Sader:2005}, be applied with great care under conditions were there is a significant contribution from non-conservative forces.

Although the authors cited in the previous section propose reasons for drawing a distinction between a traditional force and the effects arising from Pauli exclusion, the increase in kinetic energy and momentum resulting from the requirement for wavefunction antisymmetry nonetheless ultimately result in an interaction which is measured as a repulsive force in a DFM experiment. That is, the connection between the change in kinetic energy and the total energy of the tip-sample system appears to result in a measurable, and positive (i.e. repulsive), contribution to the frequency shift due to the Pauli exclusion principle. What is important to realise from the previous sections, however, is that Pauli exclusion really is not comparable to other types of interparticle interaction. In this sense it is a phenomenon which is distinct from the four fundamental forces, i.e. strong, weak, electromagnetic (in particular), and, if the graviton exists, gravity.

  \begin{figure}[t!]
  \centering
  \includegraphics[width=0.75\linewidth]{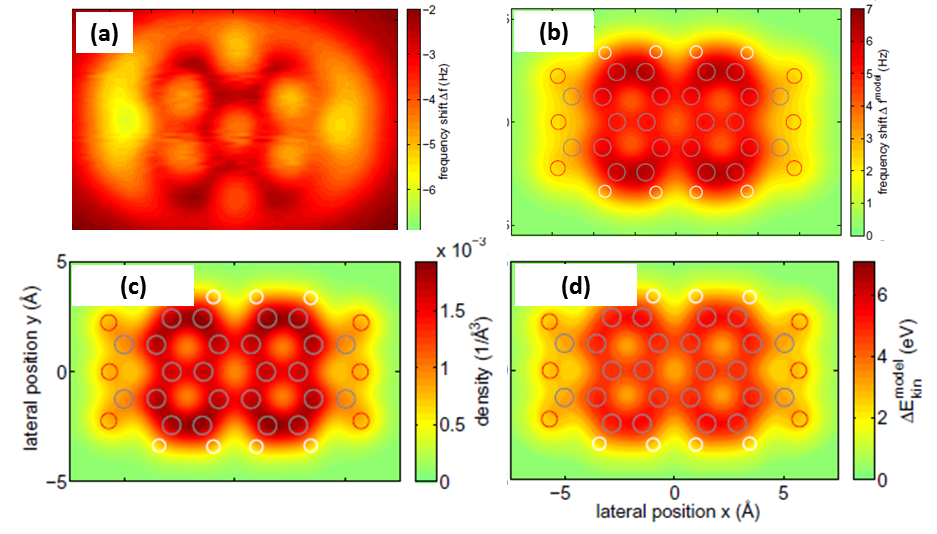}
  \caption{Comparison of \textbf{(a)} Experimental frequency shift image and \textbf{(b)} a simulated frequency shift image for a 3,4,9,10-perylenetetracarboxylic dianhydride (PTCDA) molecule calculated on the basis of the Pauli exclusion-derived change in electron kinetic energy estimated using Eqn.1.12. \textbf{(c)} Charge density of a PTCDA molecule at a given tip-sample separation. Compare with \textbf{(d)}, the change in kinetic energy at the same tip-sample separation. Figure adapted from Moll et al.\cite{Moll:2012}. \copyright Institute of Physics Publishing (2012).}
  \label{fig:Fig3}
  \end{figure}

\subsection{Intramolecular Imaging}
 Although DFM's ``sibling'' technique, scanning tunnelling microscopy (STM), has long been capable of submolecular resolution imaging, in the sense that molecular orbital density can be probed (see an earlier volume of this Springer series on Atom and Single Molecule Machines \cite{SpringerVolIII}), only DFM is capable of resolving the chemical framework or atomic structure of a molecule. This is because STM probes orbital density only within a specific energy window (set by the potential difference between the tip and sample) and in conventional tunnelling microscopy therefore only the frontier molecular orbitals are accessible\footnote{In the scanning tunnelling hydrogen microscopy (STHM)\cite{STHM1,STHM2,STHM3} variant of STM mentioned earlier, this constraint can be circumvented.}. The spatial distribution of the frontier orbital density generally does not map onto the atomic positions, and indeed often bears very little relationship to the ``ball-and-stick'' models of molecules so familiar to chemists and physicists. 
 
 As Giessibl has highlighted\cite{Giessibl:1998}, however, DFM is not restricted to probing the frontier orbital density and is sensitive to the total charge density. This is because intramolecular forces depend on the total electron density, rather than the density of states within a certain energy window\cite{Feynman:1939}. The sensitivity of DFM to the total electron density is particularly pronounced when in the Pauli exclusion regime of imaging, i.e. at very small tip-sample separations. Fig. 1.1 at the start of this chapter shows very clearly that, unlike STM, DFM in this Pauli exclusion regime produces images which are remarkably similar to the ball-and-stick structural models of molecules. 
 
On the basis of Fig. 1.3 (and related theoretical and experimental data), Moll et al.\cite{Moll:2012} argue that there is a close connection between the charge density of a molecule and the increase in electron kinetic energy due to Pauli exclusion. This assumes that \textit{(a)} the arguments regarding wavefunction curvature outlined in Sections 1.4 and 1.5 provide an accurate model of electron-electron interactions at the tip-sample junction, and \textit{(b)} the dominant effect is the change in kinetic energy, and that this can be ``deconvolved'' from the overall response of the electron density as a function of the tip-sample separation. They approximate the complicated relationship between the increase in kinetic energy and the separation of two atoms with different nuclear charges (see Eqn 6 of their paper) as follows:

\begin{equation}
\Delta E_{\textrm{\small{kin}}}(z) = A\rho_s(z)^B
\end{equation}

\noindent where $z$ is the interatomic/intermolecular separation, $\rho(z)$ is the sample charge density at separation $z$, and $A$ and $B$ are two tunable parameters. As can be seen in Fig. 1.3, this simple power law model, which involves no explicit consideration of the probe, provides good agreement with experimental frequency shift images of a 3,4,9,10-perylenetetracarboxylic dianhydride (PTCDA) molecule. We also include in Fig. 1.3, again from Moll \emph {et al.}'s paper, a comparison of the charge density of the PTCDA molecule with the increase in kinetic energy calculated using the simple model of Equation 1.11. There is again apparently good agreement, adding support to the idea that DFM is sensitive to the total charge density of the system.

What is not included in the model used to generate the simulated images in Fig. 1.3 -- although Moll and co-workers deal with this point elsewhere\cite{Gross:2012} -- is the relaxation or bending of the CO molecule at the tip apex as it is moved across the underlying PTCDA molecule. It turns out that this is an extremely important contribution to the observation of intramolecular and intermolecular contrast in DFM images and we'll return to it in the final section.

\subsection{Density depletion}
The modification of the curvature and spatial extent of the tip-sample wavefunction due to Pauli exclusion produces extensive modification to the total electron density of the system. A key aspect of this is the generation of regions of density depletion. Baerends\cite{Baerends:1992} discusses the importance of density depletion in the context of the Ag-O bond where a substantial degree of Pauli exclusion-derived depletion around the centre of the bond is observed. 

As a more recent example in the context of DFM, a number of the authors of this chapter have explored the importance of density depletion in the interpretation of images taken in the Pauli exclusion regime. The molecular system we used is that shown in Fig. 1.1(G) -- a hydrogen-bonded assembly of naphthalenetetracarboxylic diimide (NTCDI) molecules on a passivated silicon surface. Fig. 1.4 shows a comparison of the total electron density for an NTCDI assembly vs the density difference at a number of different $z$ positions of the tip above a C-C bond (Fig. 1.4(a)-(c)) and above an intermolecular region where hydrogen-bonding is expected\cite{Sweetman:2014}. Pauli exclusion results in strong tip-induced electron depletion above both the intermolecular and intramolecular bond regions.

  \begin{figure}[t!]
  \centering
  \includegraphics[width=0.75\linewidth]{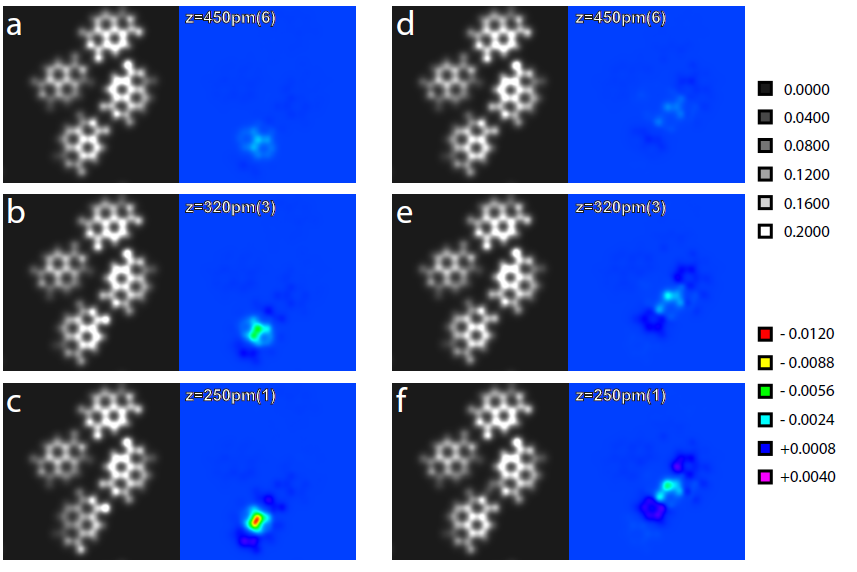}
  \caption{Total electronic density (TED) and electron density difference (EDD) calculated for an NTCDI assembly plotted 100pm above the molecular plane for a variety of different tip heights. At each tip height in a simulated $F$($z$) curve, the EDD was obtained by first calculating the TED for (i) the isolated surface, and (ii) the isolated NTCDI tip. These two densities were then summed together and
  subtracted from the relaxed total density for the full system.  The remaining quantity is the EDD. This quantifies the fraction of charge which is redistributed due to the interaction of the DFM tip and the NTCDI molecule. The TED (left) and EDD (right) are shown for an oxygen-down NTCDI probe molecule at (a)-(c) the C-C location on an NTCDI molecule, and (d)-(f) at the intermolecular H-bond location for the different tip heights specified in each figure. Figure from Sweetman et al.\cite{Sweetman:2014}. \copyright Nature Publishing Group (2014).}
  \label{fig:Fig3}
  \end{figure}

The most important insight to be derived from this analysis of density depletion is that, as is always the case in any type of scanning probe experiment (and as is well-understood across the SPM community), the influence of the tip on the imaging process must \textit{always} be carefully considered. Tip-sample interactions and convolution have been an issue for scanning tunnelling microscopy since its invention, of course, but with the advent of DFM imaging in the so-called ``Pauli regime'' the probe can certainly no longer be treated as just a perturbation of the electronic structure. The tip-sample separation for the type of high resolution images shown in Fig. 1.1 is such that the repulsive Pauli component makes a strong contribution -- the tip interacts heavily with the underlying molecule adsorbed on the sample surface. In this sense, the sample-tip apex system should be considered as one large molecule.

In the following, and final, section of this chapter we'll see just how important a role the tip can play in generating high resolution DFM images.

\section{But do we \emph{really} see bonds?} 
A key ``ingredient'' in attaining intramolecular contrast in DFM is the passivation of the tip apex. Gross \emph {et al.}\cite{Gross:2009} first showed that CO was a particularly appropriate molecule to use for imaging submolecular structure. (In the same paper, and in subsequent work\cite{GrossAPL:2013}, they compared the imaging capabilities of CO with those of other species adsorbed at the tip apex). Although deliberate functionalisation with CO is certainly not necessary to obtain intra- (and inter-)molecular contrast\cite{Sweetman:2014}, carbon monoxide remains the molecule of choice at present for high resolution DFM.

It turns out that CO is very far from a rigid probe, however, and the tilting of the molecule at the tip apex plays an essential role in the imaging process. The flexibility of CO has been studied in some detail by a number of groups \cite{Sun:2011,Welker:2012,Gross:2012,Weymouth:2014} but it is a very recent paper\cite{Hapala:2014}, available only at the condensed matter arXiv at the time of writing, on which we would like to focus here. This paper provides particularly telling insights into the extent to which the probe itself contributes to the structure seen in molecular and submolecular images.

Hapala \emph {et al.}\cite{Hapala:2014} use an exceptionally simple, but remarkably powerful, model to simulate DFM (and scanning tunnelling hydrogen microscopy\cite{STHM1,STHM2,STHM3}) images acquired either with a CO probe or any other type of tip apex. They represent the tip-sample geometry as shown in Fig. 1.5 and account for interactions between the probe and sample molecule using analytical Lennard-Jones potentials. It is very important to note that \emph{no account is taken of intra- or intermolecular charge density} in this model: the approach adopted by Hapala \emph {et al.} uses only the coordinates of the atoms within the molecule under study -- electron density due to bonding between those atoms is not incorporated in their model. In other words, the force-field does not rely on the electron density of the system. Although this might at first glance appear to be a rather crude approach (as compared to, for example, modelling the system using an \emph{ab initio} method such as density functional theory), it is nonetheless the case that their ``stripped-down'' model accurately reproduces the experimental data. This is the acid test of any theory or simulation.

  \begin{figure}[b!]
  \centering
  \includegraphics[width=0.5\linewidth]{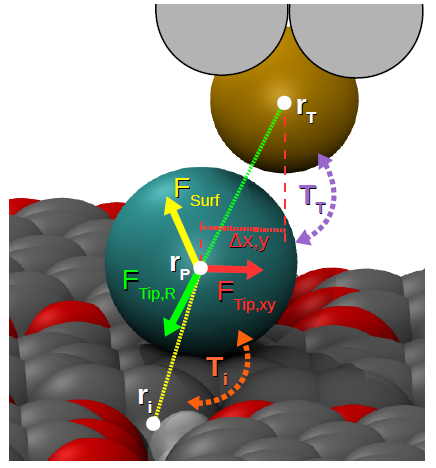}
  \caption{Schematic model of the tip-sample geometry used by Hapala et al.\cite{Hapala:2014} in their analysis of the origin of intra- and intermolecular contrast in DFM images. The final metal atom at the tip apex and the ``probe particle'' are shaded in gold and cyan respectively with the underlying molecular layer represented by the standard space-filling model. The coloured vectors show the various forces on the tip: $F_{Tip,R}$ (green) is the radial force; $F_{Tip,xy}$ (red) is the lateral force; and $F_Surf$ (yellow) is the force due to the sample molecules.  ($T_i$ and $T_t$ refer to tunnelling processes not of interest in this chapter.) Taken from Hapala\textit{ et al}.\cite{Hapala:2014}.}
  \label{fig:Fig5}
  \end{figure}

Fig. 1.6 shows a comparison between experimental DFM images and the output of Hapala \emph {et al.}'s simulations for two systems comprising assemblies of 8-hydroxyquinoline tetramers and NTCDI molecules (as discussed above in the context of Fig. 1.4), respectively. For both of these systems, intermolecular interactions are mediated by hydrogen bonding. Note, however, how the sharp intra- and intermolecular features in the simulated image of Fig. 1.6(a) agree extremely well with those in the experimental data shown in Fig. 1.6(b), despite the absence of any intra- or intermolecular charge density in the model. Fig. 1.6(c) and 6(d)
similarly show a comparison between the ``flexible tip'' model and DFM images of a hydrogen-bonded NTCDI assembly\cite{Sweetman:2014} taken by a number of the authors of this chapter. Again, intramolecular and intermolecular features are observed in the simulated image, despite the absence of any charge density due to covalent- or hydrogen-bonding. 

  \begin{figure}[t!]
  \centering
  \includegraphics[width=0.8\linewidth]{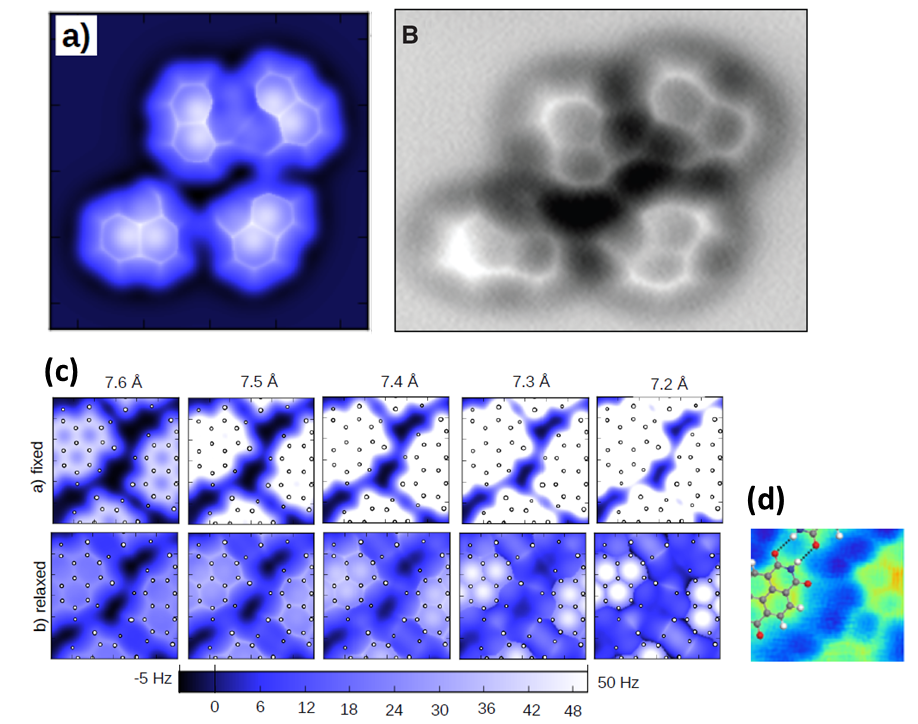}
  \caption{\textbf{(a)},\textbf{(b)}: Comparison of a simulated DFM image of a hydrogen-bonded assembly of 8-hydroxyquinoline molecules (from Hapala et al.\cite{Hapala:2014}) with the corresponding experimental DFM image taken from Zhang et al.\cite{Zhang:2013}. \textbf{(c)} Series of simulated frequency shift images at different tip-sample separations, again from Hapala et al.\cite{Hapala:2014}, of NTCDI molecules using a (top row) unrelaxed, and (bottom row) relaxed tip. \textbf{(d)} Experimental frequency shift image for comparison. (From Sweetman et al.\cite{Sweetman:2014}).}
  \label{fig:Fig6}
  \end{figure}
  
It therefore would appear that the flexibility of the probe molecule plays a major role in the imaging of intra- and intermolecular structure. But we've seen in previous sections that there's also a close correspondence between images simulated on the basis of an increase in electron kinetic energy due to Pauli exclusion and the experimental frequency shift data\cite{Moll:2012}. Moreover, the intensity of intramolecular bonds as observed by DFM is related to the Pauling bond order\cite{Gross:2012}, i.e. the charge density. Similarly, the DFM images of de Oteyza \emph {et al.}\cite{deOteyza:2013} clearly show a pronounced difference between single, double, and triple bonds. The key issue is therefore the extent to which the response of the tip to interatomic and/or intermolecular charge density is a ``first order'' vs ``second order'' contribution to the imaging mechanism, as compared to the flexibility of the probe. This is currently a very active area of debate.

In order to explore the influence of tip relaxation on the DFM images of NTCDI shown in Figs. 4 and Fig. 1.6, we (i.e. SJ, AS, LK, PJM, and co-workers\cite{Sweetman:2014}) generated simulated images using a variant of DFT where both the atomic geometry and the electronic structure of the system were ``frozen''. Despite the lack of probe relaxation, a weak feature at the expected position of the hydrogen-bond was observed. Nonetheless, another question remains: to what extent might convolution of the tip's electron density with molecular charge density at the edge of a molecule account for the observation of ``intermolecular'' features? In the supplementary information file associated with their paper, Hapala \emph {et al.}\cite{Hapala:2014} suggest that this convolution effect could be as strong as the interaction of the probe with any charge density due to an intermolecular bond. This is an exceptionally important issue which needs to be addressed in a timely fashion by the scanning probe microscopy community.

\section{Conclusions}
The history of the development of the Pauli exclusion principle provides fascinating insights into just how problematic it is to associate purely quantum mechanical concepts with classical ``real world'' analogies. In this sense, it's a shame that Pauli's \textit{Zweideutigkeit} term did not gain wider acceptance as it's a less misleading, albeit rather more prosaic, description than ``spin''.  Similarly, when we describe the Pauli exclusion principle as giving rise to a repulsive force we should bear in mind that the origin of the repulsion detected in dynamic force microscopy is not at all adequately explained via classical analogies. The interaction arises from the modification of the electrons' momentum distribution due to the increased curvature of the wavefunction imposed by the requirement for antisymmetrization in fermion systems. Classical analogies will clearly fail. Understanding the fundamental origin of the increased wavefunction curvature is ultimately, as Feynman puts it in the quote at the start of this chapter, ``\textit{deep down in relativistic quantum mechanics}".

Dynamic force microscopy provides us with direct access to the effects of Pauli exclusion on an atom-by-atom and/or molecule-by-molecule basis, and with resolution comparable to the spatial extent of a single chemical bond. This is a remarkable capability. At the time of writing it has been only five years since Gross \emph {et al.}\cite{Gross:2009} pioneered the exploitation of Pauli exclusion in force microscopy. As this variant of scanning probe microscopy is therefore in its infancy, there is potentially immense scope for detailed insights into the effects of the exclusion principle in a variety of atomic and molecular systems. However, every probe microscope image -- indeed, every image (regardless of the technique used to generate that image) -- is, at some level, a convolution of the properties of the sample and those of the imaging system. In the Pauli exclusion regime of dynamic force microscopy this convolution can be exceptionally strong. We therefore need to temper our enthusiasm for the acquisition of ultrahigh resolution images with caution regarding the interpretation of the data, as the examples included in this chapter clearly show. 

\section{Acknowledgments}
 We are very grateful for financial support from the UK Engineering and Physical Sciences Research Council in the form of a fellowship (EP/G007837), from the Leverhulme Trust (through grant F/00114/BI), from the European Commission's ICT-FET programme via the Atomic Scale and Single Molecule Logic gate Technologies (AtMol) project (www.atmol.eu), Contract No. 270028, and the ACRITAS Marie Curie Initial Training Network (www.acritas.eu). We are also very grateful for the support of the University of Nottingham High Performance Computing Facility. PJM thanks Christian Joachim (CNRS Toulouse) and Leo Gross (IBM Zurich) for helpful discussions and e-mail exchanges on the role of the exclusion principle in probe microscopy.

\bibliographystyle{unsrtnat}
\bibliography{JarvisPauli}

%Pauli, W., 1925, ¨ Uber den Zusammenhang des Abschlussees der Elektronengruppen im Atom mit der Komplexstruktur der Spektren., Zeitschrift f¨ur Physik 31, 765.
\end{document}